\documentstyle[11pt,a4,cite,epsf]{article}

\newcommand{\be}{\begin{equation}}
\newcommand{\ee}{\end{equation}}
\newcommand{\bea}{\begin{eqnarray}}
\newcommand{\eea}{\end{eqnarray}}

\begin{document}

\begin{titlepage}
\begin{flushright}
CPT-94/PE.3097\\
NORDITA - 94/62 N,P
\end{flushright}
\vspace{2cm}
\begin{center}
{\Large \bf LIGHT QUARK MASSES IN QCD}\\[1.5cm]
{\large \bf Johan Bijnens$^a$, Joaqu\'{\i}m Prades$^{a,b}$
and Eduardo de Rafael$^c$}\\[1cm]
${}^a$ NORDITA, Blegdamsvej 17,\\
DK-2100 Copenhagen \O, Denmark\\[0.4cm]
$^b$ Niels Bohr Institute, Blegdamsvej 17,\\
DK-2100 Copenhagen \O, Denmark\\[0.4cm]
$^c$  Centre  de Physique Th\'eorique\\
       CNRS-Luminy, Case 907\\
    F-13288 Marseille Cedex 9, France
\end{center}
\vspace*{1.5cm}
\begin{abstract}
We study the value of the light quark masses combination $m_u+m_d$ in
QCD using both Finite Energy Sum Rules and Laplace Sum Rules.
We have performed a detailed analysis of both the perturbative QCD and
the hadronic parametrization inputs needed in these Sum Rules.
As main result, we obtain $m_u\left(1~{\rm GeV}^2\right) + m_d
\left(1~{\rm GeV}^2\right) = \left(12\pm2.5\right)~{\rm MeV}$
for the running $\overline{MS}$ masses.
\end{abstract}
\vfill \begin{flushleft} November 1994\\
\end{flushleft} \end{titlepage}


{\bf 1.}
The values of the light quark masses in the Standard Model, are
still poorly known. We are referring to the masses of the u,d, and s
quarks which appear in the QCD Lagrangian and which explicitly break
the chiral-SU(3) flavour symmetry; the so called ``current
algebra'' quark masses. These are the parameters which, among other
things, generate the masses of
the observed low-lying SU(3) octet of pseudoscalar particles i.e.;
the Goldstone modes associated with the spontaneous breaking
of the chiral-SU(3) symmetry. The precise relation between quark
masses and pseudoscalar masses depends however on the magnitude
of the various order parameters which govern the dynamics of the
spontaneous chiral-SU(3) symmetry breaking. The order parameters are
genuinely of non-perturbative origin, and this is what causes the
difficulty in extracting the values of the light quark masses.

The masses of the $\pi$, $K$, and $\eta$ particles
are however smaller than the spontaneous
chiral symmetry breaking scale $\Lambda_{\chi}$, an empirical fact
which justifies {\it a priori} a perturbative treatment of the problem
in powers of quark masses. The appropriate field theoretical framework
to carry out this expansion  systematically is called Chiral
Perturbation Theory ($\chi PT$)\cite{W79}, \cite{GL84}, \cite{GL85}.
To lowest order in the expansion,

\bea \label {eq: pseuds.}
m^2_{\pi^+} =
m^{2}_{\pi^{0}} & = & (m_{u}+m_{d}){\bf B} + {\cal O}(m_{q}^2 ) \\
 M^{2}_{K^{+}} & = & (m_{u}+m_{s}){\bf B} +  {\cal O}(m_{q}^2 ) \\
M^{2}_{K^{0}} & = & (m_{d}+m_{s}){\bf B} +  {\cal O}(m_{q}^2 ).
\eea

\noindent
Ignoring the ${\cal O}(m_{q}^2)$ contributions, as well as
electromagnetic corrections, the expansion to lowest order
leads to the quark mass ratios estimates\cite{W78}:

\be
\frac{m_{d}-m_{u}}{m_{d}+m_{u}}=\frac{
M^2 _{K^0 }-M^2 _{K^+ }-(m^2 _{\pi^0 }-m^2_{\pi^+}
)_{\rm EM}} {m^2 _{\pi^0}}=0.29;
\ee

\noindent
and, with $\hat{m}\equiv (m_u + m_d)/2$,

\be
\frac{m_{s}-\hat{m}}{2\hat{m}}=\frac{M^2 _{K^0 }-m^2
_{\pi^0}}{m^2 _{\pi^{0}}}=12.6.
\ee

\noindent
The modification of these results due to the presence of higher order
terms as well as to electromagnetic effects, has been an active field
of research during the last few years. [For a recent review on the
subject see ref.\cite{L94a}.] The $\chi PT$ approach does not fix
however the scale of the light quark masses. For example, the
parameter {\bf B} which appears in eq. (\ref{eq: pseuds.}), is
proportional to the size of the quark-antiquark vacuum condensate in
the chiral limit. It is one of the fundamental coupling constants of
the effective Lagrangian of QCD at low-energies; but its value
cannot be determined by symmetry arguments alone. The appearance of
extra symmetries in the effective Lagrangian \cite{KM86}, preventing
the determination of the {\bf B}-like parameters, which appear
in $\chi PT$, within $\chi PT$ itself, reflects the same fact. Ideally, the
numerical simulations of lattice QCD may fix the values of these
{\bf  B}-like parameters  to a sufficiently
reliable accuracy. [For a review of progress in this
direction see e.g. ref.\cite{LATTICE93}.] Here, we want to re-examine
critically the possibility of fixing the value of the running masses
$m_{u}(s)+m_{d}(s)$ at a specific energy scale $s$, using general
analyticity properties of QCD two-point functions. We shall
then come back to a discussion of several issues which have recently
been raised in connection with the determination of the light quark
masses.

{\bf 2.}
The relevant two-point function for the determination of the
combination of quark masses $m_{u}+m_{d}$ is the one associated
with the divergence of the axial current with the quantum
numbers of the pion field (see refs. \cite{VVZNOS78} to
\cite{DdeR87},) i.e.;

\be \label {eq: two-point funct.}
\Psi_{5}(q^2)=i\int\,d^4 x e^{iq\cdot x}
  \langle 0\mid T\{\partial^{\mu}A_{\mu}^{(1-i2)}(x),
  \partial^{\nu}A^{(1+i2)}_{\nu}(0)\}\mid 0\rangle ,
\ee

\noindent
where [$\bar{q}\equiv (\bar{u},\bar{d},\bar{s})$, and $\lambda_{i}$
are Gell-Mann's flavour-SU(3) matrices,]

\be \label {eq: div. axial current}
\partial^{\mu}A_{\mu}^{(1-i2)}(x)=(m_{d}+m_{u})
  \bar{q}(x)i\gamma_{5}\frac{\lambda_{1} -i\lambda_{2}}{2}q(x)
\ee

\noindent
In QCD, the function $\Psi_{5}(q^2)$ obeys a twice subtracted
dispersion relation \newline  ($Q^2 =-q^2 \geq 0$ in our metric:
$+---$):

\be \label {eq: disp. rel.}
\Psi_{5}(q^2)=\Psi_{5}(0)+q^2 \Psi_{5}^{'}(0)+Q^4 \int_{0}^{\infty}
\frac{dt}{t^{2}}\frac{1}{t+Q^2 }\frac{1}{\pi}{\rm Im} \Psi_{5}(t).
\ee

\noindent
The subtraction constants are governed by the low-energy behaviour
of QCD. In particular

\bea \label {eq: psizero}
\Psi_{5}(0) & = & -(m_{u}+m_{d})\langle \bar{u}u+\bar{d}d \rangle
\nonumber \\
            & \equiv & 2f_{\pi}^{2} m_{\pi}^2 (1-\delta_{\pi}),
\eea

\noindent
where $\delta_{\pi}$ denotes higher order corrections. To ${\cal
O}(p^4 )$ in $\chi PT$ \cite{GL85} :

\be \label {eq: delpi}
\delta_{\pi}=4m_{\pi}^{2}f_{\pi}^2\frac{1}{{\bf f_{0}}^{4}}
(2{\bf L_{8}}-{\bf H_{2}})+{\cal O}(p^6),
\ee

\noindent
where ${\bf f_{0}}$ denotes the value of the pion coupling to vacuum
in the chiral limit, in a normalization where the corresponding
physical coupling is $f_{\pi}=$ 92.5 MeV.
The constants ${\bf L}_{8}$ and ${\bf H}_{2}$
are two specific coupling constants of the low-energy
QCD chiral effective Lagrangian of ${\cal O}(p^4)$. The other constant
$\Psi_{5}^{'}(0)$ in eq.(\ref{eq: disp. rel.}), brings in contributions
of
${\cal O}(p^6)$ so far unknown, and therefore will not be used in our
analysis.

At large Euclidean values of the momentum $Q^2 =-q^2 $, the
function $\Psi_{5}$ can be evaluated using perturbative QCD. The
calculation of the first non-trivial order (i.e., two-loops,) was
made in ref. \cite{BNdeRY81}, whereupon the first QCD analysis of the
running light quark masses were made (\cite{BNdeRY81},\cite{NdeR81},
\cite{deR81}, and \cite{DdeR87}.) The function  $\Psi_{5}$ is now
known at the three-loop level in perturbative QCD (\cite{GKL84} and
\cite{GKLS90}). Also, the present determinations of the QCD
$\Lambda_{\overline{MS}}$ parameter (see ref. \cite{RPP94} for a recent
review,) give values significantly larger than those used in the
determination made in ref.\cite{DdeR87}. These are two of the new
ingredients which we shall incorporate in the present analysis.

The basic idea to obtain the value of $m_{u}+m_{d}$ is to
use the dispersion relation in eq. (\ref{eq: disp. rel.}) to relate
the behaviour of the function
$\Psi_{5}$ obtained from perturbative QCD, including some possible
improvement from the knowledge of its leading non-perturbative
$1/Q^2$-power corrections
\cite{SVZ79} to specific integrals of the spectral function
$\frac{1}{\pi}{\rm Im}\Psi_{5}(t)$. The spectral function is an
inclusive cross-section which, {\it ideally}, could be determined in
terms of all the hadronic states with overall quantum numbers:
$I^{G}(J^{P})=1^{-}(0^{-})$, which the operator
$\partial^{\mu}A_{\mu}^{(1-i2)}(x)$ can produce from the vacuum. The
lowest hadronic state contributing to
$\frac{1}{\pi}{\rm Im}\Psi_{5}(t)$ is the pion pole. There is then a gap
until one reaches the $3\pi$-threshold. The shape of the low-energy
end of the $3\pi$-contribution can be determined from
lowest order $\chi PT$ \cite{DdeR87}. One would like however to
incorporate the contribution from higher hadronic states as well.
There are indeed states, like the $\pi(1300)$ \cite{RPP94}, with the
quantum numbers of the pion which have been observed in spectroscopic
studies of the strong interactions. These states contribute also to
the spectral function
$\frac{1}{\pi}{\rm Im}\Psi_{5}(t)$. However, as emphasized by the authors
of ref.\cite{SFK93}, their contribution  brings in, {\it a priori},
other parameters than those measured in purely hadronic reactions. This
is yet another issue which we shall also discuss in this new
analysis.


\vspace{7 mm}
{\bf 3.}
A convenient way to encode the information provided by the dispersion
relation in eq.(\ref{eq: disp. rel.}) is to use a system of Finite
Energy Sum Rules (FESR's). It has been shown that when the free
hadronic parameters in a given spectral function are constrained by
FESR's, the corresponding spectral function automatically satisfies
the ``heat evolution test''\cite{BLdeR85} of QCD-Hadronic duality.
There are three FESR's which are relevant for our purposes:

\bea \label {eq: FESR0}
\int_{0}^{s} \frac{dt}{t}\, \frac{1}{\pi} {\rm Im}\Psi_{5}(t) & = &
\frac{N_{c}}{8 \pi^{2}}[m_{u}(s)+m_{d}(s)]^{2}\,
s\left\{1+R_{0}(s)\right\} + \Psi_{5}(0);
\\ &   & \nonumber \\
\label {eq: FESR1}
\int_{0}^{s}dt\, \frac{1}{\pi} {\rm Im}\Psi_{5}(t) & = &
\frac{N_{c}}{8 \pi^{2}}[m_{u}(s)+m_{d}(s)]^{2}\, \frac{s^2}{2}
\left\{ 1+R_{1}(s)+2\frac{C_{4} \langle O_{4} \rangle}
{s^{2}} \right\}; \\
 &   &  \nonumber \\ \label {eq: FESR2}
\int_{0}^{s}dt\, t\, \frac{1}{\pi} {\rm Im}\Psi_{5}(t) & = &
\frac{N_{c}}{8 \pi^{2}}[m_{u}(s)+m_{d}(s)]^{2}\,
\frac{s^3}{3} \left\{1+R_{2}(s) - \frac{3}{2}
\frac{C_{6}\langle O_{6} \rangle}{s^3 }\right\}.
\eea

\noindent
There is yet another sum rule with two inverse powers of
$t$ in the l.h.s. integrand which we could consider, but it brings in
the unknown constant $\Psi_{5}^{'}(0)$, and therefore it is not very
useful. Other sum rules, with quadratic or higher powers of $t$,
become more and more sensitive to the behaviour of the hadronic
spectral function at high-$t$ values, where there is little
reliable experimental information at present.

The right hand sides of the three FESR's above are proportional to
$N_{c}$, the number of QCD-colours. The functions
$1+R_{n}$ with $n=0,1,2$ denote the corresponding QCD perturbative
corrections,

\be \label{eq:R1}
\frac{-1}{2 \pi i} \, \oint_s \frac{d t}{t} \left(1-\frac{t}{s}
\right) D_5 (t)
\ee
with $D_5(t)\equiv t \frac{d}{d t} \left({\Psi_5(t)}/{t}\right)$
 for $n=0$ and
\be \label{eq:R2}
\frac{-1}{2 \pi i} \, \,
 \oint_s d t \left(1-\frac{n+1}{n}\frac{t}{s}+ \frac{1}{n}\left(\frac{t}{s}
\right)^{n+1}\right) \Psi_5^{(2)}(t)
\ee

\noindent
with $\Psi_5^{(2)}(t) \equiv \frac{d^2 }{d t^2}\Psi_5(t)$ for $
n=1,2$, once
the sum of the running quark masses: $[m_{u}(s)+m_{d}(s)]^2$,
evaluated at the two-loop level and $N_c/8\pi^2$
have been factored out. This $D_5(t)$ removes
the unwanted subtraction constant
$\Psi_5'(0)$ and $\Psi_5^{(2)}$ removes
both $\Psi_5(0)$ and $\Psi_5'(0)$. Due to the
truncated expressions we have for the series of the
QCD perturbative corrections the explicit form of the
$R_{n}$-functions depends on the specific choice of the
renormalization scale
\cite{S81}; as well as on whether or not the QCD-running of the
two-point $\Psi_{5}$-function over the circle of radius
$s$ in the complex $t$-plane is taken into account\footnote{See
ref.\cite{LeDP92} for a discussion of the relevance of this dependence
in the case of the hadronic width of the $\tau$.}.
To check the numerical influence  of the truncation in the
QCD series for $\Psi_5(t)$ we shall consider
two possible choices of the $R_{n}$-functions in the present
analysis. One where we resum the leading and next-to-leading
logs after the integral in eqs. (\ref{eq:R1}) and (\ref{eq:R2})
over the circle of radius $s$  is done, by choosing the renormalization
scale at $\nu^2 = s$, and another one where
we first resum the leading and next-to-leading logs by choosing the
renormalization scale at $\nu^2 = -t$, which we shall call ``improved,''
since it takes into account the possible large QCD-running of the
two-point $\Psi_5$-function over the circle of radius $s$.
Other choices like $|\nu^2/s| \neq 1$ in the first case
or $|\nu^2/t| \neq 1$ in the second choice do not improve the
convergence of the perturbative QCD series for the
$\Psi_5$-function.

The quantities $C_{4}\langle O_{4}\rangle$ and $C_{6}\langle
O_{6}\rangle$ in the FESR's above, are a short-hand
notation for the first non-perturbative power corrections of
dimensions four and six. They are dominated respectively by the
gluon condensate: $C_{4}\langle O_{4}\rangle \simeq
 \frac{\pi}{N_c}\langle \alpha_{s} G^2 \rangle$, and the four-quark
vacuum condensate which, in the vacuum saturation approximation
\cite{SVZ79}, gives the estimate: $C_{6}\langle O_{6}\rangle \simeq
\frac{1792}{27N_c}\pi^{3}\alpha_{s}\langle \bar{q} q \rangle^{2}$, a
result rather similar to the one obtained from the leading behaviour
in the
$1/N_{c}$-expansion. In our numerical analysis we have
allowed these non-perturbative parameters to have values within a
rather generous range :

\bea \label{eq:C4O4}
C_{4}\langle O_{4}\rangle & = & (0.08\pm 0.04) {\rm GeV}^4 , \\
\label{eq:C6O6}
C_{6}\langle O_{6}\rangle & = & (0.04\pm 0.03) {\rm GeV}^6.
\eea
We use three active quark flavours in the QCD formulae and
 $\Lambda^{(3)}_{\overline{MS}} = 300\pm150$ MeV \cite{RPP94}.

The procedure we use to extract $m_{u}+m_{d}$ from the FESR's above,
is the same as the one discussed in ref.\cite{DdeR87}. First, one
fixes the choice of the upper end value of $s$ in the hadronic
integrals, by demanding a good duality between the hadronic ratio of
sum rules:

\be \label{eq: hadronic ratio}
{\cal R}_{had.}(s)\equiv \frac{3}{2s} \frac{\int_{0}^{s}dt\, t\,
\frac{1}{\pi} {\rm Im}\Psi_{5}(t)} {\int_{0}^{s}dt\, \frac{1}{\pi}
{\rm Im}\Psi_{5}(t)},
\ee

\noindent
and its QCD counterpart:

\be \label{eq: QCD ratio}
{\cal R}_{QCD}(s)\equiv \frac{1+R_{3}(s) - \frac{3}{2}
\frac{C_{6}\langle O_{6}\rangle}{s^3}} {1+R_{2}(s)+ 2
\frac{C_{4}\langle O_{4}\rangle}{s^2}}.
\ee

\noindent
With $s$ fixed within the duality region, one then solves for
$m_{u}+m_{d}$ using the second sum rule in (\ref{eq: FESR1}); and for
$\delta_{\pi}$, defined in eq.(\ref{eq: delpi}), from the first sum
rule in (\ref{eq: FESR0}).

\vspace{7 mm}
{\bf 4.}
Another technique to exploit the information encoded in
the dispersion relation in eq.(\ref{eq: disp. rel.}), consists in using
sum rules which relate moments of the Laplace transform of the hadronic
spectral function to their QCD-counterparts. Here, the master
equation is \cite{NdeR81}

\bea \label{eq:Laplace}
\frac{1}{(M^2)^2}\int_{0}^{\infty}dt\,
e^{-\frac{t}{M^2}}\,
\frac{1}{\pi}{\rm Im}\Psi_{5}(t) & = &
\frac{N_c}{8\pi^2}[m_{u}(M^2)+m_{d}(M^2)]^2  \nonumber \\
 &\times & \left\{1+\Delta(M^2)+ 2\frac{C_{4}\langle
O_{4}\rangle}{M^4}+
\frac{3}{2}\frac{C_{6}\langle  O_{6} \rangle}{M^6} \right\},
\nonumber \\
\eea

\noindent
where $\Delta(M^2)$ denotes the appropriate QCD perturbative
corrections, once the overall running mass dependence at the
$M^2$-scale has been factored out; and the leading and next to
leading non-perturbative $1/M^2$-power corrections are
explicitly shown. Here again, one has to find first a window of
duality in the $M^2$-variable. Usually, this is
fixed from the comparison of the $M^2$-dependence of the ratio of
hadronic moments:

\be \label{eq: hadronic Laplace ratio}
{\cal L}(M^2,s_{0})\equiv \frac{\int_{0}^{s_{0}}dt \, t \,
e^{-\frac{t}{M^2}}
\frac{1}{\pi}{\rm Im}\Psi_{5}(t)}{\int_{0}^{s_{0}}dt\,
e^{-\frac{t}{M^2}} \frac{1}{\pi}{\rm Im}\Psi_{5}(t)},
\ee

\noindent
to the $M^2$-dependence of its QCD-counterpart. Then, one solves for
the quark masses using eq.(\ref{eq:Laplace}), with $M^2$ fixed in the
duality region. Because of the exponential factor, the integral in the
l.h.s. of eq.(\ref{eq:Laplace}) weighs significantly the low energy
region of the hadronic spectral function. On the other hand, the
Laplace transform itself extends from $0$ to
$\infty$. In practice, this means that from a certain
$s_0$-threshold  onwards, the  hadronic spectral function will be
identified with its QCD perturbative counterpart. This threshold choice
brings in a new parameter to be fixed. As shown in ref.\cite{BLdeR85},
the appropriate choice of the onset of the perturbative continuum,
corresponds in fact to the upper end of the duality region which one
gets using the FESR's; i.e., $s_{0}\simeq s$. In the present analysis,
we shall use the Laplace transform technique only as an overall check
of the determination of the value of
$m_{u}+m_{d}$ obtained from the FESR's. Here we have only  used the
scaling at $\nu^2 = s$. The renormalization scale choice dependence
 due to the truncation of the QCD perturbative series is
 similar to the one found in the FESR's analysis.

\vspace{7mm}
{\bf 5.}
We shall now discuss the hadronic
phenomenological input for the
spectral function $\frac{1}{\pi}{\rm Im}\Psi_{5}(t)$ which appears in the
integrand in the l.h.s. of the FESR's in eqs.(\ref{eq: FESR0}),
(\ref{eq: FESR1}), and (\ref{eq: FESR2}).

As already said before, the lowest hadronic state contributing to
$\frac{1}{\pi}{\rm Im}\Psi_{5}(t)$
is the pion pole. There is then a gap until one reaches the
$3\pi$-threshold. The shape of the low-energy end of the
$3\pi$-threshold can be calculated using lowest order $\chi PT$,
with the result, [$\lambda(a,b,c)=a^2 +b^2 +c^2
-2ab-2bc-2ca$]:

\be \label{eq: HSF}
\frac{1}{\pi}{\rm Im}\Psi_{5}(t)=
2f_{\pi}^{2}m_{\pi}^{4}\delta (t-m_{\pi}^{2})+
\theta (t-9m_{\pi}^{2})
\frac{2f_{\pi}^{2}m_{\pi}^{4}}{(16\pi^{2}f_{\pi}^{2})^2}\frac{t}{18}
\rho_{\chi}^{3\pi}(t);
\ee

\noindent
with \footnote{The corresponding expression in eq.(4.14) of
ref.\cite{DdeR87} has several misprints which we have corrected
here.}

\bea \label{eq: chiral spectral}
\rho_{\chi}^{3\pi}(t) & = &
\int_{4m_{\pi}^{2}}^{(\sqrt{t}-m_{\pi})^2} \frac{du}{t}
\sqrt{1-\frac{4m_{\pi}^{2}}{u}}\left\{5+\frac{1}{2}\frac{1}
{(t-m_{\pi}^{2})^2} \left[(t-3(u-m_{\pi}^{2}))^2\right. \right.
\nonumber \\
 & +&   \left. \left. 3\lambda
(t,u,m_{\pi}^{2})(1-\frac{4m_{\pi}^{2}}{u})+20m_{\pi}^{4}\right]+
\frac{1}{(t-m_{\pi}^{2})}\left[3(u-m_{\pi}^{2})-t+9m_{\pi}^{2}
\right] \right\}.
\eea

\noindent
Notice that the chiral power counting of the pion pole term
contribution  to $\frac{1}{\pi}{\rm Im}\Psi_{5}(t)$ is ${\cal O}(p^2 )$;
while the contribution from the $\chi PT$ $3\pi$-continuum is ${\cal
O}(p^6 )$. From the point of view of the large-$N_{c}$ counting
rules, the pion pole contribution is leading; i.e., ${\cal
O}(N_{c})$, while the
$\chi PT$ $3\pi$-continuum is suppressed down to ${\cal
O}(1/N_c)$. Because of the existence of resonance states, we expect,
however, the three pion continuum to be enhanced, and contribute to
${\cal O}(N_{c})$ as well. The question is how to fold the observed
$\pi'$ resonance states to the spectral function above.
The authors of ref.\cite{SFK93} have
shown how, ideally, one could measure directly the spectral function
we are interested in from very high statistics analyses of
$\tau \rightarrow \nu_{\tau}+3\pi$ decays. In the absence of this
direct experimental information, it is important to discuss how
these $\pi '(0^{-+})$-states can contribute, and how
to propose an hadronic ansatz for
$\frac{1}{\pi}{\rm Im}\Psi_{5}(t)$ as model independent as possible.

The situation is in fact rather similar to the one already
encountered in the more familiar case of the vector-vector correlation
function with the quantum numbers of the $\rho$. In
this case, the relevant two-point function obeys a once subtracted
dispersion relation:

\be \label{eq: VDR}
\Pi(q^{2})= \Pi(0) + q^2\int_{0}^{\infty}
\frac{dt}{t} \frac{1}{t-q^2 }
\frac{1}{\pi}{\rm Im}\Pi(t),
\ee

\noindent
and to lowest order in $\chi PT$ (${\cal O}(p^4)$ in this case,) the
spectral function obtained from the $2\pi$-cut discontinuity is:

\be
\frac{1}{\pi}{\rm Im}\Pi(t)=\frac{1}{16\pi^{2}}\frac{1}{3}
(1-\frac{4m_{\pi}^{2}}{t})^{\frac{3}{2}}\:
\theta (t-4m_{\pi}^{2}).
\ee

\noindent
Again, this contribution is non-leading from the point of view of
the large-$N_{c}$ counting rules, but we know that this is not
the full story. The $\rho$ -resonance also contributes to
$\frac{1}{\pi}{\rm Im}\Pi(t)$, and its contribution is indeed leading
[${\cal O}(N_{c})$] in the large-$N_{c}$ limit. Knowing the mass
$M_{V}$ and the width $\Gamma_{V}$ (i.e., the $\rho \rightarrow 2\pi$
coupling
$g_{V}$) from strong interaction experiments, is not enough, {\it a
priori}, to fix the contribution of the same $\rho$-resonance to the
spectral function
$\frac{1}{\pi}{\rm Im}\Pi(t)$. Indeed, at the narrow width
approximation, the
$\rho$-contribution to the spectral function:

\be \label{eq: NWA}
\frac{1}{\pi}{\rm Im}\Pi(t)=f_{V}^{2}M_{V}^{2}\pi \delta(t-M_{V}^{2}),
\ee

\noindent
brings in the coupling $f_{V}$ of the $\rho$ to the external
vector-isovector current, which is different to the coupling $g_{V}$.
These two coupling constants can however be related under the rather
reasonable assumption of an unsubtracted dispersion relation for the
electromagnetic form factor of the pion, and $\rho$-dominance of
the form factor discontinuity. Then, the evaluation of the real part
of the form factor at $q^2 =0$, constrains the two couplings to
satisfy the relation \cite{EGLPdeR89}:

\be \label {eq: EGLPdeR}
1=f_{V}g_{V}\frac{M_{V}^{2}}{f_{\pi}^{2}}.
\ee

The same relation follows, if we parametrize the spectral function
$\frac{1}{\pi}{\rm Im}\Pi(t)$
with a Breit-Wigner shape function, and we fix the overall
normalization to coincide, at the $2\pi$-threshold, with the one
provided by the lowest order $\chi PT$-prediction:

\be \label{eq: BWSFP}
\frac{1}{\pi}{\rm Im}\Pi(t)=\theta (t-4m_{\pi}^{2})
\frac{1}{16\pi^{2}}\frac{1}{3}
(1-\frac{4m_{\pi}^{2}}{t})^{\frac{3}{2}}\:
\frac{(M_{V}^{2}-4m_{\pi}^{2})^{2} +
M_{V}^{2}\Gamma_{V}^{2}}{(M_{V}^{2}-t)^{2} +
M_{V}^{2}\Gamma_{V}^{2}}.
\ee

\noindent
Again, at the narrow width approximation

\be
\frac{1}{(M_{V}^{2}-t)^{2} + M_{V}^{2}\Gamma_{V}^{2}} \Rightarrow
\frac{1}{M_{V}\Gamma_{V}}\pi \delta (t-M_{V}^{2}),
\ee

\noindent
and assuming that in the limit of massless pions $\Gamma_{V}$ is
dominated by the
$2\pi$-mode,
results in the expression

\be \label {eq: BWNW}
\frac{1}{\pi}{\rm Im}\Pi(t)=\frac{f_{\pi}^{4}}{g_{V}^{2}M_{V}^{2}}.
\ee

\noindent
The identification of the two narrow width expressions
(\ref{eq: NWA}) and (\ref{eq: BWNW}) leads then to the same relation
as in eq.(\ref{eq: EGLPdeR}).

It is well known that the hadronic parametrization in eq.(\ref{eq:
BWSFP}) is in fact in very close agreement with the low-energy
phenomenological spectral function extracted from the
$e^{+}e^{-}$ annihilation cross-section data.
Following this analogy, we propose to parametrize the
$3\pi$-continuum contribution to $\frac{1}{\pi}{\rm Im}\Psi_{5}(t)$,
modifying eq.(\ref{eq: chiral spectral}) as follows:

\be \label{eq: rho continuum}
\rho_{\chi}^{3\pi}(t) \Rightarrow \rho_{had.}(t),
\ee

\noindent
with

\bea \label{eq: hadronic continuum}
\rho_{had.}(t) & = & \left|
F(M_{1},\Gamma_{1};M_{2},\Gamma_{2};\xi;t)\right| ^2
\int_{4m_{\pi}^{2}}^{(\sqrt{t}-m_{\pi})^2} \frac{du}{t}
\sqrt{1-\frac{4m_{\pi}^{2}}{u}} \nonumber \\
 &\times & \left\{5+\frac{1}{2}\frac{1} {(t-m_{\pi}^{2})^2}
\left[(t-3(u-m_{\pi}^{2}))^2+20m_{\pi}^4\right. \right. \nonumber \\
 &+ & \left. 3\lambda
(t,u,m_{\pi}^{2})(1-\frac{4m_{\pi}^{2}}{u})\, \frac{(M_{\rho}^2
-4m_{\pi}^2)^2 +M_{\rho}^2 \Gamma_{\rho}^2}{(M_{\rho}^2 -u)^2
+M_{\rho}^2 \Gamma_{\rho}^2}\right] \nonumber \\
 &+ & \left.
\frac{1}{(t-m_{\pi}^{2})}\left[3(u-m_{\pi}^{2})-t+9m_{\pi}^{2}
\right]\right\}.
\eea
The function F which appears as an overall factor in the r.h.s. encodes
the modulation due to the presence of two
$\pi{'}$-resonances\footnote{A possible, not yet confirmed, new
$\pi'$-resonance
has been reported in \cite{newpi'}. However its inclusion in the present
analysis does not affect any of the results.}
with possible interference ($\xi$ is a complex
parameter,) and with masses
$M_{1,2}$ and widths
$\Gamma_{1,2}$:

\be \label{eq: piprimes}
F[M_{1},\Gamma_{1};M_{2},\Gamma_{2};\xi;t]=
\frac{\left| \frac{1}{(t-M_{1}^2)+i\Gamma_{1}M_{1}}+\xi
\frac{1}{(t-M_{2}^2)+i\Gamma_{2}M_{2}}\right|^2}
{\left| \frac{1}{(9m_{\pi}^2-M_{1}^2)+i\Gamma_{1}M_{1}}+\xi
\frac{1}{(9m_{\pi}^2-M_{2}^2)+i\Gamma_{2}M_{2}}\right| ^2}.
\ee

\noindent
In the hadronic parametrization above we
have also allowed for a possible modulation of the $(1^{-})$
$2\pi$-subchannel in the
$3\pi$-continuum to take into account the effect of the
$\rho(770)$-resonance. The factor $F$ has been normalized to 1 at the
3$\pi$-threshold, but we shall also discuss the effect of possibly larger
normalization values as suggested in ref. \cite{SFK93}.

\begin{figure}[hbt]
\epsfxsize=11cm
\hspace{2cm}\epsfbox{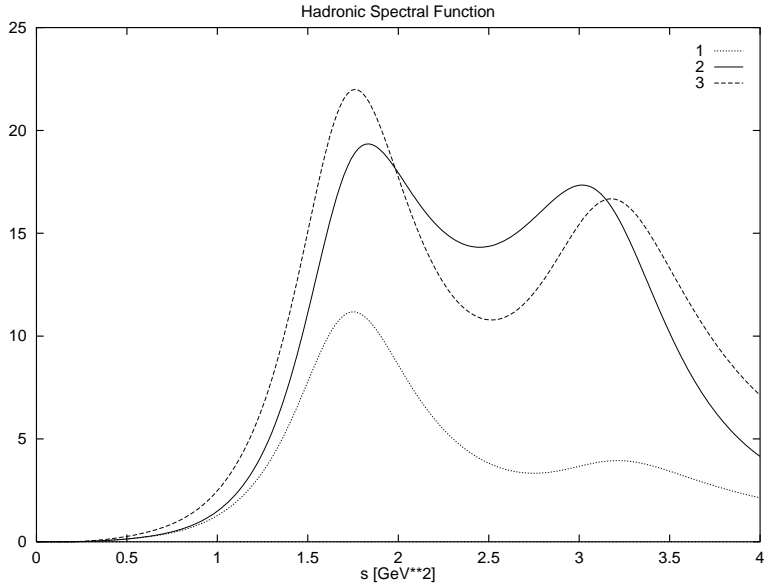}
\caption{ Shape of the hadronic continuum contribution to the spectral
function in eq. (\protect\ref{eq: HSF})
in units of $10^{-6}~{\rm GeV}^4$. The
continuous curve is the one which gives the best duality
with QCD.  The dotted curve is the one which
reproduces the best fit to the experimental curves which observe the
$\pi'(1770)$ in hadronic interactions. The dashed curve is the one with
the best duality when the normalization at the 3$\pi$-threshold
is twice as large as in eq. (\protect{\ref{eq: piprimes}}).}
\label{fig: HC}
\end{figure}

\vspace{7mm}
{\bf 6.}
The shape of the hadronic continuum contribution to the spectral
function in (\ref{eq: HSF}) i.e., the function

\be
\left. \frac{1}{\pi}{\rm Im}\Psi_{5}(t)\right|_{\rm{HC}}\, \equiv \,
\theta (t-9m_{\pi}^{2})
\frac{2f_{\pi}^{2}m_{\pi}^{4}}{(16\pi^{2}f_{\pi}^{2})^2}\frac{t}{18}
\rho_{had.}(t);
\ee
\noindent
has been plotted in Fig.\ref{fig: HC} for $s$-values between
the 3$\pi$-threshold and 4 GeV$^2$. The masses and widths of the
$\pi'$-resonances have been fixed to the estimated values in
ref.\cite{RPP94} (i.e., in MeV units:)

\be
M_{1}=1300\pm 100,\; \Gamma_{1}=400\pm200; \; \; M_{2}=1770\pm 30,
\;\Gamma_{2}=310\pm 50.
\ee

\noindent
The three curves in Fig.\ref{fig: HC} correspond to the
$\xi$-parameter
values: $\xi=0.234+i\,0.1$ (the dotted curve); $\xi=-0.23+i\,0.65$
(the continuous curve)
and $\xi=0.4+i\,0.4$ (the dashed curve) where the
normalization at the 3$\pi$-threshold
is twice as large as the one in  eq. (\ref{eq: piprimes}).
The first $\xi$-value (dotted curve) is the one
which reproduces the best fit to the experimental curves which
observe the $\pi'(1770)$ in hadronic interactions \cite{B84}. The
second $\xi$-value (continuous curve) is the one which gives the best
duality with QCD.
The third $\xi$-value is one which gives a good duality with QCD with a
normalization at
the 3$\pi$-threshold equal to 2. Other normalizations lead to
similar results for the quark masses
after imposing the QCD-Hadronic duality constraint.

\begin{figure}
\epsfxsize=11cm
\hspace{2cm}\epsfbox{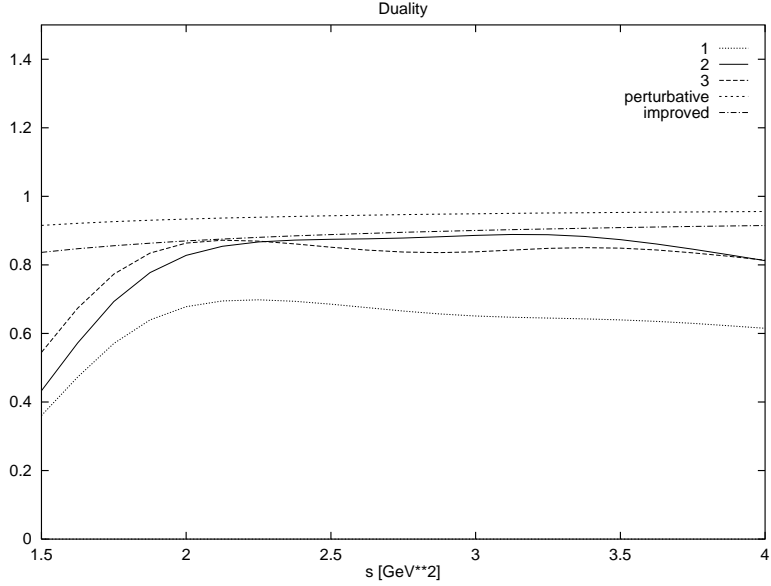}
\caption{ QCD-Hadronic duality test. The short-dashed curve is the one
corresponding to the QCD ratio ${\cal R}_{QCD}(s)$ in
eq. (\protect\ref{eq: QCD ratio}) with $\nu^2 = s$. The
dash-dotted curve is the ``improved'' ratio.
The other curves are the ones corresponding
to the hadronic ratio
${\cal R}_{had}(s)$ in eq. (\protect\ref{eq: hadronic ratio}) using
the same input values as in Fig. \protect \ref{fig: HC}.}
\label{fig: HDQCD}
\end{figure}

The duality test is shown in Fig.\ref{fig: HDQCD}.
Here, the short dashed curve is the one corresponding to the QCD ratio
${\cal R}_{QCD}(s)$ in eq.(\ref{eq: QCD ratio})
and the dash-dotted one is the same ratio evaluated with the
``improved'' set of $R_{1,2}$-functions discussed above.
Most of the difference between the perturbative and the
``improved'' version is in $R_2$ and very little in $R_1$. This explains
why the quark masses obtained show less variation than the duality ratio.
The other curves are the hadronic ratios, with the same input as in Fig.
\ref{fig: HC}. As can be seen curve 1 has no good duality but 2 and 3 are
acceptable. We show it further
as an illustration of the type of variation obtained
with the hadronic input.

\begin{figure}
\epsfxsize=11cm
\hspace{2cm}\epsfbox{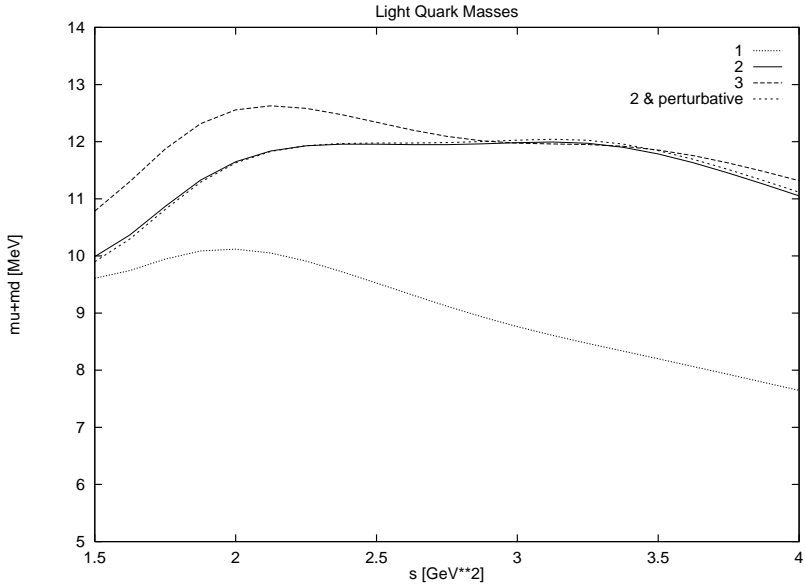}
\caption{The running quark masses $m_{u}(1{\rm GeV}^2)+m_{d}(1
{\rm GeV}^2)$ in
the
$\overline {MS}$-scheme. The three curves correspond to the results we
obtain  solving for
$m_{u}+m_{d}$ in the FESR in eq. (\protect\ref{eq: FESR1}), using the
``improved'' expression for the QCD-function $R_{1}(s)$,
and using the three hadronic parametrizations shown in
Fig.\protect \ref{fig: HC}. The short-dashed curve shows the change if we
use for the hadronic input the 2nd parametrization and the perturbative
expression for the QCD counterpart.}
\label{fig: quark masses}
\end{figure}

Figure \ref{fig: quark masses} shows the results we get for the quark
masses. Here we show the value we get for the sum of running
masses $m_{u}(1{\rm GeV}^2)+m_{d}(1{\rm GeV}^2)$ in MeV in the
$\overline {MS}$-scheme. The curves correspond to the result we
obtain  solving for
$m_{u}+m_{d}$ in the FESR in eq.(\ref{eq: FESR1}), using the
``improved'' expression for the QCD-function $R_{1}(s)$,
and using the three hadronic parametrizations shown in
Fig.\ref{fig: HC}. The continuous curve corresponds to the hadronic
parametrization which passed best the QCD-Hadronic duality test.
For this case we also show the result with the perturbative expression for
$R_1$.
The
stability of the result is rather remarkable. If the same good
QCD-Hadronic duality is wanted, then a
modification of the hadronic parametrization of the 3$\pi$-continuum
in (\ref{eq: hadronic continuum}),
including, for instance, a global normalization factor
at the 3$\pi$-threshold to the two $\pi'$-resonances overall function $F$
in (\ref{eq: piprimes}) (as suggested in ref. \cite{SFK93}),
or varying the complex
parameter $\xi$, leads to a spectral function very similar to the
continuous curve in Fig. \ref{fig: HC} and therefore to the same
results from the FESR's.
As an example we have shown the effect of a
normalization factor of 2 and a variation of $\xi$.
Thus, we quote the results for the best duality hadronic
parametrization:

\be
\label{eq:masses}
m_u (1 {\rm GeV}^2) + m_d (1 {\rm GeV}^2) = (12 \pm 2.5) ~{\rm MeV},
\ee
where the error reflects changes in the input parameters.

\vspace{7mm}
{\bf 7.}
It is well known that
the scalar channels can be affected by instanton effects, however,
there is no clear computational scheme to get a
reliable estimate of these effects (for an attempt see \cite{GN93}).
Nevertheless, it is also obvious from the results in this reference
that if the  duality
region is for energies around $s \simeq (2.5 \sim 3.5)$ GeV$^2$ as we have,
these effects can be absorbed in the quoted error bars for the
masses above from the FESR  analysis.

\begin{figure}
\epsfxsize=11cm
\hspace{2cm}\epsfbox{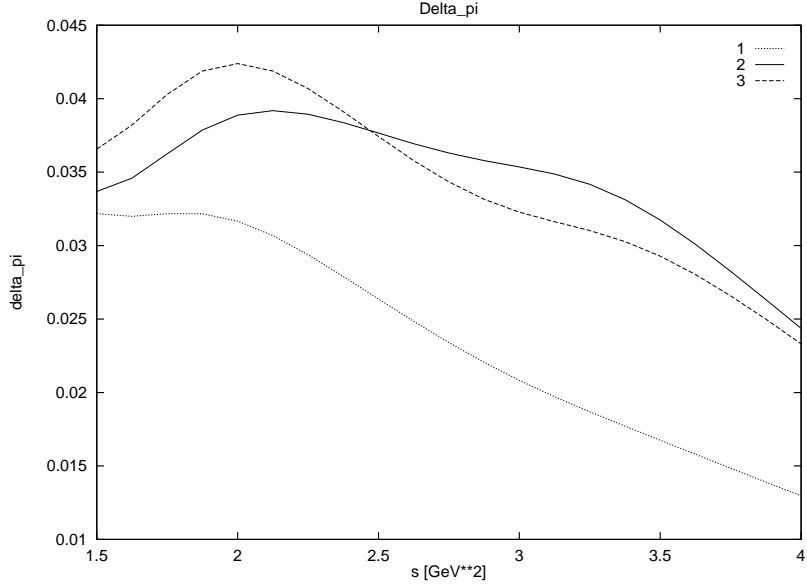}
\caption{The parameter $\delta_{\pi}$ defined in
eq.(\protect\ref{eq: psizero}) which results from solving for
$\delta_{\pi}$ in the first FESR in eq.(\protect\ref{eq:
FESR0}). The three curves correspond to the hadronic parametrizations
shown in Fig.\protect\ref{fig: HC} and using the perturbative QCD expression
for $R_0$. }
\label{fig: delta pi}
\end{figure}

Figure \ref{fig: delta pi} shows the result we get for the parameter
$\delta_{\pi}$ defined in eq.(\ref{eq: psizero}). This is the result
of solving for $\delta_{\pi}$ in the first FESR in (\ref{eq:
FESR0}). Again the continuous curve is the one corresponding to the
hadronic parametrization which passed best the QCD-Hadronic duality
test i.e., the one which leads to the continuous lines also in
Figs.\ref{fig: HC} and \ref{fig: quark masses}. For the QCD function
$R_0$ we use the perturbative expression.
The improved one is within the errors. The stability of
the result is less impressive than for the quark masses in Fig.
\ref{fig: quark masses}, but good enough (notice the vertical
scale,) to obtain an estimate of this parameter.

\be
\label{eq: deltapi}
\delta_\pi = (3.5 \pm 1) \% .
\ee

\noindent
{}From this result and using (\ref{eq: delpi}) we can obtain
for the scale independent quantity \cite{GL85}
$2{\bf L_{8}}-{\bf H_{2}}$
\be
\label{2l8-h2}
2{\bf L_{8}}-{\bf H_{2}} = (2.9\pm1.0)\times 10^{-3} \, ,
\ee
and from (\ref{eq: psizero})
\be
\label{eq:quark condensate}
\frac{\langle \bar u u + \bar d d \rangle (1 {\rm GeV}^2)}
{2} =  -(0.013\pm 0.003) {\rm GeV}^3 .
\ee
This result is for the non-normal ordered quark condensate (see
refs.  \cite{JM94} and \cite{CDPK94}.)

We can also combine our result for the light quark masses
in (\ref{eq:masses}) with the recent results for the strange quark mass
in refs. \cite{JM94,CDPK94} and \cite{LATTICE93} to get the ratio of quark
masses (we use $m_s(1~{\rm GeV}^2) = (175\pm25)~{\rm MeV}$):

\be
\label{eq: ratio}
r\equiv \frac{m_s}{\hat m} = 29 \pm 7 .
\ee
Although the error bars are large, this result is free of
the uncertainties noted in \cite{KM86} since it is obtained from
QCD.

We can also combine our result with the estimation of the
next-to-leading corrections to Dashen's theorem in refs.
\cite{BIJ93} and \cite{DHW93}

\be
\left(M_{K^+}^2-M_{K^0}^2\right)_{\rm EM} =
(1.9\pm 0.4) \,
\left(m_{\pi^+}^2-m_{\pi^0}^2\right)_{\rm EM} .
\ee
This result translates into, using the ratio from \cite{GL85},
\be
\frac{m_d-m_u}{m_d +m_u} =
\frac{m_\pi^2}{M_K^2}
\frac{\left(M_{K^0}^2-M_{K^+}^2\right)_{\rm QCD}}{M_K^2 - m_\pi^2}
\frac{m_s^2-\hat{m}^2}{4 \hat{m}^2}
= (0.52 \pm 0.05) \times 10^{-3}
\, \left(r^2 -1 \right)
\ee
and with the values for the quark mass ratio $r$ above
we get
\be
\frac{m_u}{m_d}=0.44\pm 0.22 .
\ee

\begin{figure}
\epsfxsize=11cm
\hspace{2cm}\epsfbox{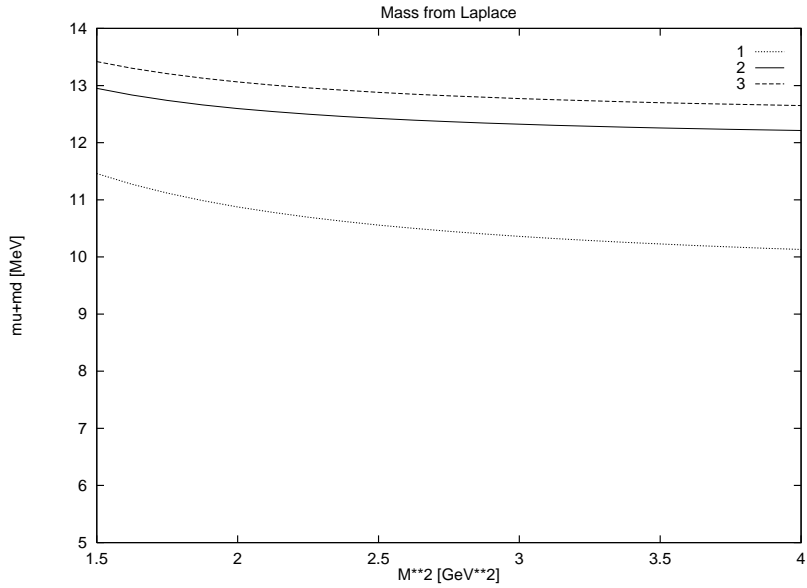}
\caption{The running quark masses $m_{u}(1{\rm GeV}^2)+m_{d}(1
{\rm GeV}^2)$ in the
$\overline {MS}$-scheme, obtained with the Laplace transform sum rule
in eq.(\protect \ref{eq: Laplace quark masses}). Curves 1,2 and 3
correspond to the hadronic parametrizations shown
in Fig.\protect \ref{fig: HC} and for $s = 2.5~{\rm GeV}^2$.
The results for $s=3.5~{\rm GeV}^2$ are very similar.}
\label{fig: Laplace masses}
\end{figure}

We shall next discuss the consistency of our results  for
the quark masses, using FESR's,  with the results from the
Laplace transform technique. With the replacement in eq. (\ref{eq: rho
continuum}) incorporated in the spectral function in (\ref{eq: HSF}),
the master equation in (\ref{eq:Laplace}) can be written as

\bea
\label{eq: Laplace quark masses}
\frac{2f_{\pi}^2m_{\pi}^4}{M^4}\left\{ e^{-\frac{m_{\pi}^2}{M^2}} +
\frac{1}{(16\pi^2 f_{\pi}^2)^2}
\frac{1}{18}\int_{9m_{\pi}^2}^{s} dt \,
e^{-\frac{t}{M^2}} \, \rho_{had}(t) \right\} =
\frac{N_{c}}{8\pi^2}[m_{u}(s)+m_{d}(s)]^2 \nonumber \\
\times \,
\left\{1+P(M^2,s) +2 \frac{C_4\langle O_4\rangle}{M^4}
+ \frac{3}{2}\frac{C_6\langle O_6 \rangle}{M^6}\right\},
\eea
\noindent
with $P(M^2,s)$ the corresponding perturbative QCD
expression given in the appendix. Solving for the quark
masses in this expression results in the plots shown in
Fig. \ref{fig: Laplace masses}. As in the previous figures we show
the three hadronic
parametrizations of Fig. \ref{fig: HC} for $s=2.5~{\rm GeV}^2$.
The results for $s=3.5~{\rm GeV}^2$ are very similar.
The agreement between the two
determinations is very good. We have also checked that the hadronic
parametrization gives a reasonable duality for the ratio in eq.
(\ref{eq: hadronic Laplace ratio}).

\section*{Acknowledgements}
EdeR thanks NORDITA where part of this work was done for hospitality.
JP thanks
the Leon Rosenfeld foundation (K{\o}benhavns Universitet) for support
and CICYT (Spain) for partial support under Grant No AEN-93-0234.
This work was also done under NorFA grant No. 93.15.078/00
and INTAS 93-283.
\appendix
 \def\theequation{A.\arabic{equation}}
\setcounter{equation}{0}


\section*{Appendix}

\vspace{7mm}
\noindent{\bf 1.}
{\sc The Two-Point Function $\Psi_{5}(q^2)$ in perturbation theory.}
\vspace{7mm}

With $\nu^2$ the renormalization-scale, and $Q^2=-q^2$:
\bea \label{eq:psi5qcd}
\left. \Psi_{5}(q^2)\right|_{QCD} & = & \frac{N_{c}}{8\pi^2}
  [m_{u}(\nu^{2})+m_{d}(\nu^{2})]^2 \, Q^2 \,
\left\{ b_{0}\log\frac{Q^2}{\nu^2}+a_{0}  \right. \nonumber \\
 &+& \frac{\alpha_{s}(\nu^2)}{\pi}\left[c_{1}\log^2 \frac{Q^2}{\nu^2}+
b_{1}\log\frac{Q^2}{\nu^2}+a_{1}\right] \nonumber \\
 & +& \left(\frac{\alpha_{s}(\nu^2)}{\pi}\right)^2 \left[d_{2}\log^3
\frac{Q^2}{\nu^2}+ c_{2}\log^2\frac{Q^2}{\nu^2}+
b_{2}\log\frac{Q^2}{\nu^2}+a_{2}\right] \nonumber \\
 & +&  \left.
{\cal O} \left[\left(\frac{\alpha_{s}}{\pi}\right)^3\right]
 \right\}.
\eea
The non-trivial coefficients are
$b_{0}  =  1$; $b_{1}  =  {17}/{3}$ and
\be
b_{2}  =  \frac{10801}{144}-\frac{39}{2}\zeta (3)
            -\left(\frac{65}{24}-\frac{2}{3}\zeta (3)\right)
         n_{f},
\ee
with $n_f$ the number of active quark flavours.
The other coefficients are fixed by renormalization group
properties with the results:
$c_{1}  =  -\gamma_{1}/2$;
$c_{2}  =
  (b_{1}/4)(\beta_{1}-2\gamma_{1})-
            \gamma_{2}/2$ and
$d_{2}  =
            (-\gamma_{1}/12)(\beta_{1}-2\gamma_{1})$,
where $\beta_{i}$ and $\gamma_{i}$ are the coefficients of the
perturbation theory series expansion in powers of
$\alpha_{s}(\nu^2)/{\pi}$ of the $\beta(\alpha_{s})$-function
associated with the QCD coupling constant renormalization:
$\beta_{1}  =  -11/2+n_{f}/3$ and
$\beta_{2}  =  -{51}/{4}+{19}n_{f}/12$;
and the $\gamma(\alpha_{s})$-function associate with mass
renormalization:
$\gamma_{1}  =  2$ and
$\gamma_{2}  =  {101}/{12}-{5}n_{f}/18$.  From eq. (\ref{eq:psi5qcd})
we can easily obtain the expressions for $D_5(t)$ and $\Psi^{(2)}_5(t)$
with $t=q^2$, needed in eqs. (\ref{eq:R1}) and (\ref{eq:R2})
for the evaluation of the FESR's.

\vspace{7 mm}
\noindent{\bf 2.}
{\sc The Spectral Function $\frac{1}{\pi}{\rm Im}\Psi_{5}(t)$ in
perturbation theory.}
\vspace{7 mm}

\bea
\left. \frac{1}{\pi}{\rm Im}\Psi_{5}(t)
\right|_{{\rm QCD}} &= & \frac{N_{c}}{8\pi^2}
[m_{u}(\nu^2)+m_{d}(\nu^2)]^2 \, t \, \left\{1+
\frac{\alpha_{s}(\nu^2)}{\pi}\left[b_{1}+2c_{1}\log\frac{t}{\nu^2}
\right] \right.  \nonumber \\
 & +& \left. \left(\frac{\alpha_{s}(\nu^2)}{\pi}\right)^2 \left[
b_{2}+2c_{2}\log\frac{t}{\nu^2}+d_{2}(3\log^2
\frac{t}{\nu^2}-\pi^2)\right] \right. \nonumber \\
&+& \left. {\cal O} \left[\left(\frac{\alpha_{s}}{\pi}\right)^3\right]
\right\}.
\eea

\vspace{7 mm}
\noindent{\bf 3.}
{\sc The running coupling constant and the running mass.}
\vspace{7mm}

With
$a_{1}\equiv 2/\left(-\beta_{1}\log\frac{Q^2}{\Lambda^2}
\right)$,
the running coupling at the two-loop level, at the scale $Q^2$, is
\be
a(Q^2)\equiv \frac{\alpha_{s}(Q^2)}{\pi}=
a_{1}\left\{ 1-a_{1}\frac{\beta_{2}}{\beta_{1}}
\log \log \frac{Q^2}{\Lambda^2} \right\}.
\ee

At that level, the renormalization group summed expression for the
running mass at the $Q^2$-scale is:
\be
m(Q^2)=\overline{m} \, \left[ a(Q^2)\right]
^{\frac{\gamma_{1}}{-\beta_{1}}}\;
\left[1+\frac{\beta_{2}}{\beta_{1}}a(Q^2)
\right]^{\left(\frac{\gamma_{1}}{\beta_{1}}-
\frac{\gamma_{2}}{\beta_{2}}\right)},
\ee
where $\overline{m}$ fixes the arbitrary constant of motion of the
renormalization flow differential equation, which defines an invariant
mass.

\vspace{7 mm}
\noindent{\bf 4.}
{\sc The Laplace Transform ${\cal M}_{5}(M^2)$ in Perturbation
Theory.}
\vspace{7mm}

The operator $\hat{L}$ which transforms
$\Psi_{5}^{(2)}(Q^2)$ into the function
\be
\label{eq:M5}
{\cal M}_{5}(M^2)=\frac{1}{(M^2)^3}\int_{0}^{\infty}dt\,
e^{-\frac{t}{M^2}}\; \frac{1}{\pi}{\rm Im} \Psi_{5}(t),
\ee
is
\be
\hat{L}\equiv  \lim_{
N\rightarrow \infty
}
\;\left \{ \frac{(-Q^2)^{N}}{\Gamma(N)}
\frac{\partial^{N}}{(\partial Q^2)^{N}}
\right\}
\ee
with $M^2 \equiv Q^2/N$ finite.

As explained in the text, in practice  we have identified
the hadronic spectral function $\frac{1}{\pi}
{\rm Im} \Psi_5(t)$ in the r.h.s. of eq. (\ref{eq:M5}) with
its perturbative QCD expression
from some threshold $s_0 \simeq s$ onwards. This allows us to
rewrite  eq. (\ref{eq:Laplace}) as
\bea
\frac{1}{(M^2)^2}\int_{0}^{s}dt\,
e^{-\frac{t}{M^2}}\; \frac{1}{\pi}{\rm Im} \Psi_{5}(t)
&=& \frac{N_c}{8 \pi^2} [m_u(s)+m_d(s)]^2 \nonumber \\
&\times&  \left\{
1+P(M^2,s)+ 2 \frac{C_4 \langle O_4 \rangle}{M^4}+\frac{3}{2}
\frac{C_6 \langle O_6 \rangle}{M^6} \right\}
\eea
with $P(M^2,s)$ defined by
\bea
\frac{1}{(M^2)^2}\int_{0}^{s}dt\,
e^{-\frac{t}{M^2}}\;
\left. \frac{1}{\pi}{\rm Im} \Psi_{5}(t)\right|_{QCD}
&=& \frac{N_c}{8 \pi^2} [m_u(s)+m_d(s)]^2
  \left\{1+P(M^2,s)\right\} .
\eea


\newpage
\section*{Figures}

\end{document}